\documentclass[referee]{chjaa}           

\usepackage{graphicx}                   
\usepackage{epsfig}

\begin{document}

   \title{The relation between black hole masses and Lorentz factors of the jet components in blazars}

   \volnopage{Vol.0 (200x) No.0, 000--000}      
   \setcounter{page}{1}          

   \author{Ming Zhou
      \inst{1,2,3}\mailto{}
   \and Xinwu Cao
      \inst{1,2\star}}
   \offprints{Ming Zhou}                   

   \institute{Shanghai Astronomical Observatory, CAS, Shanghai 200030, China\\
             \email{mzhou@shao.ac.cn; cxw@shao.ac.cn}
        \and
         Joint Institute for Galaxy and Cosmology (JOINGC) of SHAO and USMC, 80 Randan Road, Shanghai 200030, China
       \and
        Graduate School of the Chinese Academy of Sciences, BeiJing 100039, China}

   \date{Received~~ month day; accepted~~ ~~month day}

\abstract{We explore the relation between black hole mass ($M_{\rm
BH}$) and the motion of the jet components for a sample of blazars.
The Very Long Baseline Array (VLBA) 2cm Survey and its continuation:
Monitoring of Jets in active galactic nuclei (AGNs) with VLBA
Experiments (MOJAVE) have observed 278 radio-loud AGNs, of which 146
blazars have reliable measurements on their apparent velocities of
jet components.  We calculate the minimal Lorentz factors for these
sources from their measured apparent velocities, and their black
hole masses are estimated with their broad-line widths.  A
significant intrinsic correlation is found between black hole masses
and the minimal Lorentz factors of the jet components, which the
Eddington ratio is only weakly correlated with the minimal Lorentz
factor, which may imply that the Blandford-Znajek (BZ) mechanism may
dominate over the Blandford-Payne (BP) mechanism for the jet
acceleration (at least) in blazars.
   }

   \authorrunning{Ming Zhou \and Xinwu Cao }            
   \titlerunning{Black hole masses and Lorentz factors of the jet components}  

\maketitle

%
%
\section{Introduction}           
\label{sect:intro}
Relativistic jets have been observed in many radio-loud AGNs, which
are believed to be formed very close to the black holes. The
currently most favored models of the jet formation are BZ and BP
mechanisms (Blandford \& Znajek 1977; Blandford \& Payne 1982). In
these mechanisms, the power of jet is extracted from the disk or
black hole rotational energy. The disk-jet connection has been
investigated by many authors in different ways (Rawlings \& Saunders
1991; Falcke \& Biermann 1995;  Cao \& Jiang 1999; 2001; 2002, Xie
et al. 2007; Xie et al. 2008).

Some different approaches were proposed to estimate the masses of
the black holes in AGNs, such as the gas kinematics near a black
hole (see Ho \& Kormendy 2000 for a review and references therein).
The central black hole mass derived from the direct measurements on
the gases moving near the hole is reliable, but unfortunately, it is
available only for very few AGNs. For most AGNs, the velocities of
the clouds in broad line regions (BLR) can be inferred from the
widths of their broad emission lines. If the radius of the BLR is
available, the mass of the central black hole can be derived from
the broad-line width on the assumption that the clouds in the BLR
are gravitationally bound and orbiting with Keplerian velocities
(Dibai 1980). The radius of the BLR can be measured by using the
reverberation-mapping method from the time delay between the
continuum and line variations (Peterson 1993; Netzer \& Peterson
1997). Long-term monitoring on the source is necessary for applying
this method to derive the radius of the BLR, which leads to a small
amount of AGNs with measured black hole masses in this way.
Alternatively, a tight correlation was found between the size of the
BLR and the optical continuum luminosity, which can be used to
estimate the size of the BLR in an AGN from its optical luminosity
and then the black hole mass (e.g., Wandel, Peterson \& Malkan 1999;
Kaspi et al. 1996; 2000; Laor 2000).

The kinematic properties of
the jet components in blazars were revealed by multi-epoch VLBI observations (e.g.,
Kellermann et al. 2004; Lister et al. 2005).
In this paper, we use a large sample of blazars, of which the proper motions were well
measured with VLBA, to explore the relations between the jet speeds and physical properties
of blazars, i.e., the black hole masses and Eddington ratios.

The cosmological parameters $\Omega_{\rm M}=0.3$,
$\Omega_{\Lambda}=0.7$, and $H_0=70~ {\rm km~s^{-1}~Mpc^{-1}}$ have
been adopted in this work.


\section{Sample}
\label{sect:data}
We start with a sample of radio-loud quasars and BL Lac objects with
measured apparent velocities of jet components. The sample is
compiled by searching the literature to include all blazars with
available proper motion data of the jets. Most data are taken from
several surveys, such as, the original flux-limited MOJAVE-I sample,
and the extended MOJAVE-II sample
\footnote[1]{http://www.physics.purdue.edu/astro/MOJAVE}. We find
that 278 sources have multi-epoch VLBI observations, of which 146
blazars have reliable measured apparent velocities. Their black hole
masses are estimated with the broad-line widths and
broad-line/continuum luminosities, which leads to 78 sources with
measured black hole masses.

\section{Black Hole Masses and Minimal Lorentz Factors}
\label{sect:analysis}

In order to estimate their $M_{\rm BH}$, we search the literatures
for all the available measurements of the full width at half maximum
(FWHM) for broad-lines H$\alpha$, H$\beta$, Mg {\sc ii}, C {\sc iv}
or Ly$\alpha$ lines, as well as the fluxes of these lines. For the
sources without line flux data, we adopt their continuum fluxes
instead. We find that one source in our sample has very narrow
broad-lines (FWHM$<$ 1000 km~s$^{-1}$), which is similar to typical
narrow lines. It should be cautious on the black hole mass estimates
for this source, since we cannot rule out the possibility that this
line may be the narrow component emitted from the narrow line region
(e.g., Gu et al. 2001). We therefore rule out this source. For most
BL Lac objects, their broad-line emissions are too weak to be
measured, and we estimate their $M_{\rm BH}$ with the empirical
relation between $M_{\rm BH}$ and bulge luminosity $L_{\rm bulge}$.
We list all the data of the sample in Table 1.  Columns (1)-(2)
represent the source's IAU name and redshift, respectively. The
lines used to estimate $M_{\rm BH}$ from their luminosity and the
references are listed in Columns (3) and (4), respectively. In
Columns (5), we list the lines, of which the widths are used to
estimate the $M_{\rm BH}$. We list the estimated $M_{\rm BH}$ in
Columns (7). The data of the BL Lac objects in this sample are
summarized in Table 2.

For blazars, the optical/UV continuum may be contaminated by the
beamed synchrotron emission from the jets. Wu et al. (2004) compared
the black hole masses obtained for a sample of radio-loud quasars
with both the line and continuum, and they found that the masses
obtained with line luminosity are systematically lower that those
obtained with continuum. In this work, the black hole masses $M_{\rm
BH}$ are estimated by using the line width of either one of these
lines: Mg {\sc ii}, H$\beta$, or H$\alpha$, and the line
luminosities (or the optical/UV continuum, if the line luminosity is
unavailable). McGill et al. (2008) analyzed a sample of 19 AGNs of
which all three lines were observed in optical wavebands, and they
obtained a set of 30 internally self-consistent recipes for
estimating $M_{\rm BH}$ from a variety of observables with different
intrinsic scatters. Whenever more than one recipes are available for
estimating the black hole mass, we always choose the one with the
minimal intrinsic scatter (see McGill et al. 2008 for the details).
We therefore use the broad-line emission instead of the optical/UV
continuum to estimate $M_{\rm BH}$, provided their line luminosities
are available. When more than one empirical correlations are
applicable, we use one of following relations in order:
\begin{equation}
 \log M_{\rm BH}=6.384+2\log \left(\rm FWHM_{\rm Mg_{\ II}} \over 1000 \mbox{\rm km s}^{-1} \right)+0.55\log \left({\it L}_{\rm H\alpha}\over10^{44}\mbox{\rm erg s}^{-1}\right)
 \end{equation}
\begin{equation}
 \log M_{\rm BH}=6.711+2\log \left(\rm FWHM_{\rm Mg_{\ II}} \over 1000 \mbox{\rm km s}^{-1} \right)+0.56\log \left({\it L}_{\rm H\beta}\over10^{44}\mbox{\rm erg s}^{-1}\right)
 \end{equation}
\begin{equation}
 \log M_{\rm BH}=6.711+2\log \left(\rm FWHM_{\rm Mg_{\ II}} \over 1000 \mbox{\rm km s}^{-1} \right)+0.56\log \left({\it L}^{'}_{\rm H\alpha}\over10^{44}\mbox{\rm erg
 s}^{-1}\right),
 \end{equation}
\begin{equation}
 \log M_{\rm BH}=6.930+2\log \left(\rm FWHM_{\rm H\alpha} \over 1000 \mbox{\rm km s}^{-1} \right)+0.56\log \left({\it L}_{\rm H\beta}\over10^{44}\mbox{\rm erg
 s}^{-1}\right),
 \end{equation}
\begin{equation}
 \log M_{\rm BH}=6.747+2\log \left(\rm FWHM_{\rm H\beta} \over 1000 \mbox{\rm km s}^{-1} \right)+0.55\log \left({\it L}_{\rm H\beta}\over10^{44}\mbox{\rm erg
 s}^{-1}\right),
 \end{equation}
\begin{equation}
 \log M_{\rm BH}=6.420+2\log \left(\rm FWHM_{\rm H\beta} \over 1000 \mbox{\rm km s}^{-1} \right)+0.56\log \left({\it L}_{\rm H\alpha}\over10^{44}\mbox{\rm erg
 s}^{-1}\right),
 \end{equation}
\begin{equation}
 \log M_{\rm BH}=6.747+2\log \left(\rm FWHM_{\rm H\beta} \over 1000 \mbox{\rm km s}^{-1} \right) +0.55\log \left({\it L}^{'}_{\rm H\beta}\over10^{44}\mbox{\rm
 erg,
 s}^{-1}\right),
 \end{equation}
\begin{equation}
 \log M_{\rm BH}=6.747+2\log \left(\rm FWHM_{\rm C_{\ IV}} \over 1000 \mbox{\rm km s}^{-1} \right) +0.55\log \left({\it L}^{'}_{\rm H\beta}\over10^{44}\mbox{\rm erg s}^{-1}\right)+\log
 0.5,
 \end{equation}
\begin{equation}
 \log M_{\rm BH}=6.990+2\log \left(\rm FWHM_{\rm Mg_{\ II}} \over 1000 \mbox{\rm km s}^{-1} \right)+0.518\log \left({\it L}_{\rm 5100}\over10^{42}\mbox{\rm erg
 s}^{-1}\right),
 \end{equation}
\begin{equation}
 \log M_{\rm BH}=7.026+2\log \left(\rm FWHM_{\rm H\alpha} \over 1000 \mbox{\rm km s}^{-1} \right)+0.518\log \left({\it L}_{\rm 5100}\over10^{42}\mbox{\rm erg
 s}^{-1}\right),
 \end{equation}
\begin{equation}
 \log M_{\rm BH}=7.026+2\log \left(\rm FWHM_{C_{\ IV}} \over 1000 \mbox{\rm km s}^{-1} \right)+0.518\log \left({\it L}_{\rm 5100}\over10^{42}\mbox{\rm erg s}^{-1}\right)+\log
 0.5,
 \end{equation}
where $ M_{\rm BH}$ is in units of $M_{\odot}$, $ L^{'}_{\rm
H\alpha}$ and $L^{'}_{\rm H\beta}$ are estimate form $L_{\rm
Ly\alpha}$, $L_{\rm C_{\ IV}}$ or $L_{\rm Mg {\ II}}$ by their
relative rations (Gaskell, Shields \& Wampler 1981; Francis et al.
1991),and $L_{5100}$ is nuclear luminosity $\lambda L_{\lambda}$ at
$\lambda =\rm 5100\rm \AA$.

We also estimate $M_{\rm BH}$ using Eqs. (9), (10), and (11) for the
black holes with continuum luminosities, of which the masses can
also be estimated with line luminosities. We compare the black hole
masses with these two different methods in Fig. 1.  It is indeed
found that the masses estimated with line luminosities are
systematically lower that those estimated with continuum
luminosities, which is consistent with Wu et al. (2004)'s
conclusion.

For the BL Lac objects in this sample, we use the empirical relation
between host galaxy absolute magnitude at R-band $M_{\rm R}$ and
$M_{\rm BH}$ proposed by Bettoni et al. (2003),
\begin {equation}
\log M_{\rm BH}=-0.50 M_{\rm R}-3.00,
\end {equation}
to estimate their black hole masses.

Although the apparent velocities of the jet components were measured
by VLBI observations, the intrinsic speeds of the jet components are
still unavailable, as the viewing angles of the jets are unknown for
most sources in this sample. However, we can derive the minimal
Lorentz factors from the observed apparent velocities of jet
components using:
\begin {equation}
\gamma_{\rm min}=(1+\beta_{\rm app} ^{2})^{0.5},
\end {equation}
and then analyze their relations with other physical quantities of
the sources. For the sources with more than one measured moving
component, we always select the one moving fastest, as we intend to
explore the acceleration mechanism of the jets in blazars (see Cohen
et al. 2007 for the detailed discussion).

\section{Results}
\label{sect:analysis}

In Fig. 2, we plot the relation between black hole masses $M_{\rm
BH}$ and the minimal Lorentz factors $\gamma_{\rm min}$ of the jets.
The linear regression gives
\begin{equation}
\log \gamma_{\rm min}=  0.31 \log M_{\rm BH}-1.80
\end{equation}

A significant correlation is found between these two quantities at
99.6 per cent confidence (Spearman rank correlation analysis), and
the correlation coefficient is 0.33. It should be noted with caution
that this correlation may be caused by the common dependence of
redshift. In Fig. 3, we plot the relation between redshift $z$ and
the minimal Lorentz factor $\gamma_{\rm min}$, and only a weak
correlation is found at 93.6 per cent confidence between these two
quantities. We perform the Spearman partial rank correlation
analysis (Macklin 1982), and we find that the partial correlation
coefficient is 0.27 after subtracting the common redshift
dependence. The significance of the partial rank correlation is
2.39, which is equivalent to the deviation from a unit variance
normal distribution if there is no correlation present (see Macklin
1982 for the details). A summary of the results of partial rank
correlation analysis is listed in Table 3.

We also perform a correlation analysis on the sources in the
restricted redshift range $0.1<z< 2.1$. For this subsample of 72
sources, a correlation at 97.6 per cent confidence is still present
between $\gamma_{\rm min}$ and $M_{\rm BH}$, while almost no
correlation between $\gamma_{\rm min}$ and z is found (at 55.2 per
cent confidence). It appears that the correlation between
$\gamma_{\rm min}$ and $M_{\rm BH}$ is an intrinsic one, not caused
by the common redshift dependence.

The bolometric luminosity ($L_{\rm bol}$) is estimated by assuming
$L_{\rm bol}\approx10L_{\rm BLR}$ (e.g., Liu et al. 2006). For some
sources without measured broad-line luminosities, we estimate the
bolometric luminosities from the optical continuum luminosities
using the relation of $L_{\rm bol}\approx 9\lambda L_{\rm
\lambda,opt}$ ($\lambda =\rm 5100\rm \AA$) (Kaspi et al. 2000). We
plot the relation between the Eddington ratio ($L_{\rm bol}/L_{\rm
Edd}$) and $\gamma_{\rm min}$ of the jets in Fig. 5. The linear
regression gives
\begin{equation}
\log \gamma_{\rm min}=0.11 \log L_{\rm bol}/L_{\rm Edd}+0.87.
\end{equation}
We find that only a weak correlation between $L_{\rm bol}/L_{\rm
Edd}$ and $\gamma_{\rm min}$ (at 93.5 per cent confidence) is
present.

\section{Discussion}

We find an intrinsic correlation between black hole masses and the
minimal Lorentz factors of jet components for a sample of blazars,
while no significant correlation between the Eddington rations and
the Lorentz factors is present for the same sample. Our main
statistical results will not be altered, even if those black holes
with masses estimated with continuum luminosities are removed. Our
statistical results provide useful clues to the mechanisms of jet
formation and acceleration in blazars.

It is believed that the growth of massive black holes in the centers
of galaxies is dominantly governed by mass accretion in AGN phases
(e.g., Soltan 1982; Yu \& Tremaine 2002). The massive black holes
will be spun up through accretion, as the black holes acquire mass
and angular momentum simultaneously though accretion. The spins of
massive black holes may also be affected by the mergers of black
holes. A rapidly rotating new black hole will be present after the
merger of two black holes, only if the binary's larger member
already spins quickly and the merger with the smaller hole is
consistently near prograde, or if the binary's mass ratio approaches
unity (Hughes \& Blandford 2003). The comoving space density for
heavier black holes is much lower than that for lighter black holes
(e.g., see the black hole mass function in Yu \& Tremaine 2002),
which means that the probability of the mergers of two black holes
with similar masses is lower for heavier black holes. This implies
the spins of heavier black holes are mainly regulated by accretion
rather than the mergers. Thus, it is natural to expect (in
statistical sense) that the heavier black holes have higher spin
parameters $a$ than their lower mass counterparts. Volonteri et al.
(2007) studied on how the accretion from a warped disc influences
the evolution of black hole spins and concluded that within the
cosmological framework, one indeed expects most supermassive black
holes in elliptical galaxies to have on average higher spin than
black holes in spiral galaxies, where random, small accretion
episodes (e.g., tidally disrupted stars, accretion of molecular
clouds) might have played a more important role. The jets can be
accelerated to higher speeds by the heavier black holes, because
they are spinning more rapidly (Blandford \& Znajek 1977). The
intrinsic correlation between black hole masses and the minimal
Lorentz factors of jet components found in this work is consistent
with the Blandford-Znajek mechanism. The properties of accretion
disks accretion disk are related with the dimensionless accretion
rates $\dot{m}$ ($\dot{m}=\dot{M}/\dot{M}_{\rm Edd}\propto L_{\rm
bol}/L_{\rm Edd}$). No significant correlation between $L_{\rm
bol}/L_{\rm Edd}$ and $\gamma_{\rm min}$ is found, which implies
that the jet acceleration may not be related with the properties of
the accretion disk, which may imply that the jet formation is not
sensitive to the disk structure. This is, of course, quite puzzling,
and to be verified by the future work with a larger blazar sample.
Our statistical results implies that the BZ mechanism may dominate
over BP mechanism for the jet acceleration in blazars.

\begin{acknowledgements}
We thank the referee for the helpful comments/suggestions, and
'MOJAVE survey' for sharing their data on the website. This work is
supported by the NSFC (10773020), and the CAS (grant KJCX2-YW-T03).
This research has made use of the NASA/IPAC Extragalactic Database
(NED), which is operated by the Jet Propulsion Laboratory,
California Institute of Technology, under contract with the National
Aeronautics and Space Administration.
\end{acknowledgements}

\clearpage
\begin{table*}
\begin{center}
\begin{small}
\caption{\label{lines} The data for quasars. Col. (1): IAU source
name. Col. (2): Redshift. Col. (3): log of the minimal Lorentz
factor of jet. Col. (4):  the lines used to estimate the black hole
masses from their fluxes. Col. (5): references for Col. (4). Col.
(6): the lines used to estimate the black hole masses from their
FWHMs. Col. (7): references for Col. (6).  Col. (8): the black hole
masses.}
\begin{tabular}{lcccccccc}
\hline\hline
Source&z &$\log\gamma_{\rm min}$&line &Refs. &line &Refs. &$\log M_{\rm BH}/M_{\odot}$\\
(1)& (2)& (3)&(4)&(5)&(6)&(7)&(8)\\
\hline
   0016$+$731  &    1.781    &  0.760    & Mg {\sc ii}  & L96   & $Ly\alpha$ C {\sc iv} Mg {\sc ii}   & L96   &  8.93    \\
   0035$+$413  &    1.353    &  0.926    & Mg {\sc ii}  & SK93  & Mg {\sc ii}  & SK93  &  8.53    \\
   0106$+$013  &    2.107    &  1.461    & C {\sc iv}   & B89   & C {\sc iv}   & B89   &  8.83    \\
   0112$-$017  &    1.365    &  0.159    & Mg {\sc ii}  & B89   & C {\sc iv}   & B89   &  7.85    \\
   0119$+$041  &    0.637    &  0.291    &  H$\beta$  & JB91b     &  H$\beta$  & JB91a     &  8.50    \\
   0133$+$476  &    0.859    &  0.350    & Mg {\sc ii}  & L96   &  H$\beta$  & L96   &  8.30    \\
   0212$+$735  &    2.367    &  1.071    & Mg {\sc ii}  & L96   & $Ly\alpha$ C {\sc iv} Mg {\sc ii}   & L96   &  8.48    \\
   0333$+$321  &    1.263    &  1.030    & Mg {\sc ii}  & B94   & Mg {\sc ii}  & S91   &  8.49    \\
   0336$-$019  &    0.852    &  1.006    & Mg {\sc ii}  & B89   &  H$\beta$  & JB91a     &  8.78    \\
   0403$-$132  &    0.571    &  1.279    &  H$\beta$  & M96   &  H$\beta$  & S97   &  8.77    \\
   0420$-$014  &    0.915    &  0.933    & Mg {\sc ii}  & B89   & Mg {\sc ii}  & S97   &  8.84    \\
   0440$-$003  &    0.844    &  0.183    & Mg {\sc ii}  & B89   &  H$\beta$  & JB91a     &  8.63    \\
   0605$-$085  &    0.872    &  1.367    & Mg {\sc ii}  & S93   & Mg {\sc ii}  & S93   &  8.43    \\
   0607$-$157  &    0.324    &  0.047    &  H$\beta$  & H78   &  H$\beta$  & H78   &  7.63    \\
   0736$+$017  &    0.191    &  1.082    &  H$\beta$  & B96   &  H$\beta$  & S97   &  8.23    \\
   0738$+$313  &    0.630    &  0.892    &  H$\beta$  & B96   &  H$\beta$  & JB91a     &  9.08    \\
   0804$+$499  &    1.432    &  0.552    & Mg {\sc ii}  & L96   & C {\sc iv} Mg {\sc ii}  & L96   &  8.57    \\
   0836$+$710  &    2.180    &  1.550    & Mg {\sc ii}  & L96   & $Ly\alpha$ C {\sc iv} Mg {\sc ii}   & L96   &  9.49    \\
   0850$+$581  &    1.322    &  0.903    & Mg {\sc ii}  & L96   & Mg {\sc ii}  & L96   &  9.67    \\
   0859$-$140  &    1.339    &  1.214    & Mg {\sc ii}  & B94   & Mg {\sc ii}  & S91   &  8.87    \\
   0906$+$015  &    1.018    &  1.288    & Mg {\sc ii}  & B89   & Mg {\sc ii}  & S97   &  8.63    \\
   0923$+$392  &    0.698    &  0.729    & Mg {\sc ii}  & L96   &  H$\beta$  & L96   &  9.27    \\
   0945$+$408  &    1.252    &  1.230    & Mg {\sc ii}  & L96   & C {\sc iv} Mg {\sc ii}  & L96   &  9.71    \\
   0953$+$254  &    0.712    &  1.063    &  H$\beta$  & JB91b     &  H$\beta$  & JB91a     &  8.73    \\
   1038$+$064  &    1.265    &  0.848    & Mg {\sc ii}  & B94   & Mg {\sc ii}  & S91   &  8.76    \\
   1055$+$018  &    0.888    &  0.398    & Mg {\sc ii}  & B89   & Mg {\sc ii}  & S97   &  8.45    \\
   1226$+$023  &    0.158    &  1.118    & Mg {\sc ii}  & B89   & H$\alpha$     & JB91a     &  8.76    \\
   1253$-$055  &    0.538    &  0.953    & Mg {\sc ii}  & W95   & H$\alpha$     & N79   &  8.53    \\
   1302$-$102  &    0.278    &  0.744    & Mg {\sc ii}  & B89   &  H$\beta$  & M96   &  7.90    \\
   1334$-$127  &    0.539    &  1.247    & Mg {\sc ii}  & S93   & Mg {\sc ii}  & S93   &  8.36    \\
   1458$+$718  &    0.904    &  0.833    & Mg {\sc ii}  & L96   &  H$\beta$  & L96   &  8.84    \\
   1502$+$106  &    1.839    &  1.249    & Mg {\sc ii}  &W86    & C {\sc iv}   & S97   &  8.86    \\
   1504$-$166  &    0.876    &  0.608    & Mg {\sc ii}  & H78   & Mg {\sc ii}  & H78   &  8.84    \\
   1510$-$089  &    0.360    &  1.133    & Mg {\sc ii}  &W86    & H$\alpha$     & N79   &  8.22    \\
   1532$+$016  &    1.420    &  1.147    & Mg {\sc ii}  & B89   & C {\sc iv} Mg {\sc ii}  & S97   &  8.73    \\

\hline
\end{tabular}
\end{small}
\end{center}
\end{table*}
\addtocounter{table}{-1}

\clearpage
\begin{table*}
\begin{center}
\begin{small}
\caption{Continued...}
\begin{tabular}{lcccccccc}
\hline\hline
Source&z &$\log \gamma_{\rm min}$&line &Refs. &line &Refs. &$\log M_{\rm BH}/M_{\odot}$\\
(1)& (2)& (3)&(4)&(5)&(6)&(7)&(8)\\
\hline
   1546$+$027  &    0.412    &  1.071    & Mg {\sc ii}  & B89   & H$\beta$  & S97   &  8.82    \\
   1611$+$343  &    1.401    &  1.197    & H$\beta$  & N95   & $Ly\alpha$ C {\sc iv}    & W95   &  9.49    \\
   1633$+$382  &    1.807    &  1.380    & Mg {\sc ii}  & L96   & $Ly\alpha$ C {\sc iv} Mg {\sc ii}   & L96   &  10.14   \\
   1637$+$574  &    0.751    &  1.118    & Mg {\sc ii}  & L96   & H$\beta$  & L96   &  8.68    \\
   1641$+$399  &    0.594    &  1.275    & Mg {\sc ii}  & L96   & H$\beta$  & L96   &  9.03    \\
   1642$+$690  &    0.751    &  1.222    & Mg {\sc ii}  & L96   & Mg {\sc ii}  & L96   &  8.49    \\
   1656$+$053  &    0.879    &  0.655    & H$\beta$  & B96   & Mg {\sc ii}  & S97   &  9.09    \\
   1739$+$522  &    1.379    &  0.961    & C {\sc iv}   & L96   & C {\sc iv}   & L96   &  8.20    \\
   1741$-$038  &    1.057    &  0.827    & Mg {\sc ii}  & S89   & Mg {\sc ii}  & S89   &  8.67    \\
   1828$+$487  &    0.692    &  1.110    & Mg {\sc ii}  & L96   & H$\beta$  & L96   &  8.66    \\
   1921$-$293  &    0.352    &  0.637    & H$\beta$  & JB91b     & H$\alpha$     & JB91a     &  8.38    \\
   1928$+$738  &    0.303    &  0.913    & H$\alpha$     & L96   & H$\beta$  & L96   &  8.76    \\
   2113$+$293  &    1.514    &  0.303    & Mg {\sc ii}  & S93   & Mg {\sc ii}  & S93   &  8.74    \\
   2121$+$053  &    1.941    &  1.165    & Mg {\sc ii}  & B94   & Mg {\sc ii}  & S91   &  8.60    \\
   2128$-$123  &    0.501    &  0.864    & H$\alpha$     & O02   & H$\beta$  & T93   &  9.16    \\
   2134$+$004  &    1.932    &  0.391    & C {\sc iv}   & B89   & C {\sc iv}   & B89   &  8.50    \\
   2145$+$067  &    0.999    &  0.407    & Mg {\sc ii}  & B94   & Mg {\sc ii}  & S91   &  8.61    \\
   2155$-$152  &    0.672    &  0.523    & Mg {\sc ii}  & S89   & H$\beta$  & S89   &  7.81    \\
   2201$+$315  &    0.298    &  0.848    & Mg {\sc ii}  & W95   & H$\alpha$     & JB91a     &  8.91    \\
   2216$-$038  &    0.901    &  0.748    & Mg {\sc ii}  & B94   & $Ly\alpha$ C {\sc iv} Mg {\sc ii}   & W95 S91   &  8.89    \\
   2223$-$052  &    1.404    &  1.249    & C {\sc iv}   & W95   & $Ly\alpha$ C {\sc iv} Mg {\sc ii}   & W95 S97   &  8.54    \\
   2230$+$114  &    1.037    &  0.950    & C {\sc iv}   & W95   & $Ly\alpha$ C {\sc iv} Mg {\sc ii}   & W95 S97   &  8.64    \\
   2251$+$158  &    0.859    &  1.187    & H$\beta$  & N95   & H$\beta$  & JB91a     &  8.87    \\
   2345$-$167  &    0.576    &  1.145    & H$\beta$  & JB91b     & H$\beta$  & JB91a     &  8.59    \\
   2351$+$456  &    1.986    &  1.452    & Mg {\sc ii}  & L96   & Mg {\sc ii}  & L96   &  9.22    \\
   0458$-$020  &    2.291    &  1.179    & C {\sc iv}   & B89   & $m_{\rm B}$   & ...   &  9.27    \\
   0730$+$504  &    0.720    &  1.236    & Mg {\sc ii}  & H97   & $m_{\rm B}$   & ...   &  8.84    \\
   0748$+$126  &    0.889    &  1.317    & Mg {\sc ii}  & W86   & $m_{\rm B}$   & ...   &  8.84    \\
   1012$+$232  &    0.565    &  1.012    & H$\beta$  & B96   & $m_{\rm B}$   & ...   &  8.69    \\
   1127$-$145  &    1.187    &  1.133    & Mg {\sc ii}  & W86   & $m_{\rm B}$   & ...   &  9.18    \\
   1145$-$071  &    1.342    &  0.433    & C {\sc iv}   & W86   & $m_{\rm B}$   & ...   &  8.61    \\
   1156$+$295  &    0.729    &  1.290    & H$\beta$  & B96   & $m_{\rm B}$   & ...   &  9.19    \\
   1508$-$055  &    1.191    &  1.269    & Mg {\sc ii}  & W86   & $m_{\rm B}$   & ...   &  8.97    \\
   1655$+$077  &    0.621    &  1.048    & Mg {\sc ii}  & W86   & $m_{\rm B}$   & ...   &  7.91    \\
   1726$+$455  &    0.714    &  0.602    & Mg {\sc ii}  & H97   & $m_{\rm B}$   & ...   &  8.59    \\
   1901$+$319  &    0.635    &  0.455    & H$\beta$  & G94   & $m_{\rm B}$   & ...   &  8.80    \\
   2008$-$159  &    1.178    &  0.536    & Mg {\sc ii}  & W86   & $m_{\rm B}$   & ...   &  9.56    \\
   2227$-$088  &    1.562    &  0.925    & C {\sc iv}   & W86   & $m_{\rm B}$   & ...   &  8.85    \\

\hline
\end{tabular}
\end{small}
\end{center}
\vskip 1mm $^{*}$the sources without line luminosities, we use their
optical continuum luminosities at B band to estimate $M_{\rm BH}$.
\end{table*}
\clearpage
\begin{table*}
\begin{center}
\begin{small}
\begin{tabular}{lcccccccc}
\hline
\end{tabular}
\end{small}
\end{center}
\vskip 1mm References:B89: Baldwin et al. (1989). B94: Brotherton et
al. (1996). B96: Brotherton (1996). G94: Gelderman et al. (1994).
H78: Hunstead et al. (1978). H97: Henstock et al. (1997) JB91a:
Jackson \&Browne (1991). JB91b: Jackson \& Browne (1991). L96:
Lawrence et al. M96: Marziani et al. (1996). N79: Neugebauer et al.
(1979). N95: Nerzer et al. (1995). O02: Oshlack et al. (2002). S89:
Stickel et al. (1989). S91: Steidel et al. (1991). SK93: Stickel \&
K\"uhr (1993). S93: Stickel et al. (1993). S97: Scarpa et al. (1997)
T93: Tadhunter et al. (1993). W95: Wills et al. (1995). W86: Wills
et al. (1986).
\end{table*}
\clearpage
\begin{table*}
\begin{center}
\begin{small}
\caption{$\gamma_{\rm min}$ and $M_{\rm BH}$ for BL Lac. Col. (1):
IAU source name. Col. (2): Redshift. Col. (3): log of the minimal
Lorentz factor of jet. Col. (4): absolute R host galaxy magnitude
and the references, respectively. Col. (6): the black hole masses.}
\begin{tabular}{lcccccccc}
\hline\hline
Source&z &$\log \gamma_{\rm min}$ & $M_{\rm R}(host)$ & Refs. &$\log M_{\rm BH}/M_{\odot}$\\
(1)& (2)& (3)&(4)&(5)& (6)\\
\hline
0829$+$046   &  0.180    &  1.037    &  -22.98   & U00   &  8.49    \\
1749$+$096   &  0.320    &  0.923    &  -22.68   & U00   &  8.34    \\
1807$+$698   &  0.051    &  0.480    &  -23.18   & U00   &  8.59    \\
2007$+$777   &  0.342    &  0.114    &  -22.96   & U00   &  8.48    \\
2200$+$420   &  0.069    &  0.813    &  -22.84   & U00   &  8.42    \\

\hline
\end{tabular}
\end{small}
\end{center}
\vskip 1mm References:U00: Urry et al. (2000).
\end{table*}

\clearpage
\begin{table*}
\begin{center}
\begin{small}
  \caption{The Spearman partial rank correlation analysis of the sample.
Here $r_{\rm AB}$ is the rank correlation coefficient of the two
variables, and $r_{\rm AB,C}$ the partial rank correlation
coefficient. The significance of the partial rank correlation is
equivalent to the deviation from a unit variance normal distribution
if there is no correlation present. }
\begin{tabular}{lcccccccc}
\hline\hline
 Sample&N&Correlated variables: A,B & Variable: C & $r_{\rm AB}$
& $r_{\rm AB,C}$ & significance \\ \hline
 All & 78 &$ M_{\rm BH}$,$\gamma_{\rm min}$& z &0.33 &0.27 & 2.39\\
     &    & z,$\gamma_{\rm min}$ &$M_{\rm BH}$ &0.21 &0.07 & 0.61\\
     &    & $M_{\rm BH}$, z & $\gamma_{\rm min}$&0.46 &0.42 & 3.85\\
     &    & $L_{\rm bol}/L_{\rm Edd}$,$\gamma_{\rm min}$& z & 0.21 &
     0.18& 1.52\\
 Within &72& $M_{\rm BH}$,$\gamma_{\rm min}$&  z &0.27 &0.25 & 2.14\\
  $0.1<z<2.1$   &    & z ,$\gamma_{\rm min}$ &$M_{\rm BH}$ &0.09 &-0.02 & -0.17\\
     &    & $M_{\rm BH}$, z & $\gamma_{\rm min}$&0.41 &0.40 & 3.47\\

\hline
\end{tabular}
\end{small}
\end{center}
\end{table*}
\clearpage
\begin{figure}
   \vspace{2mm}
   \begin{center}
   \hspace{3mm}\epsfig{figure=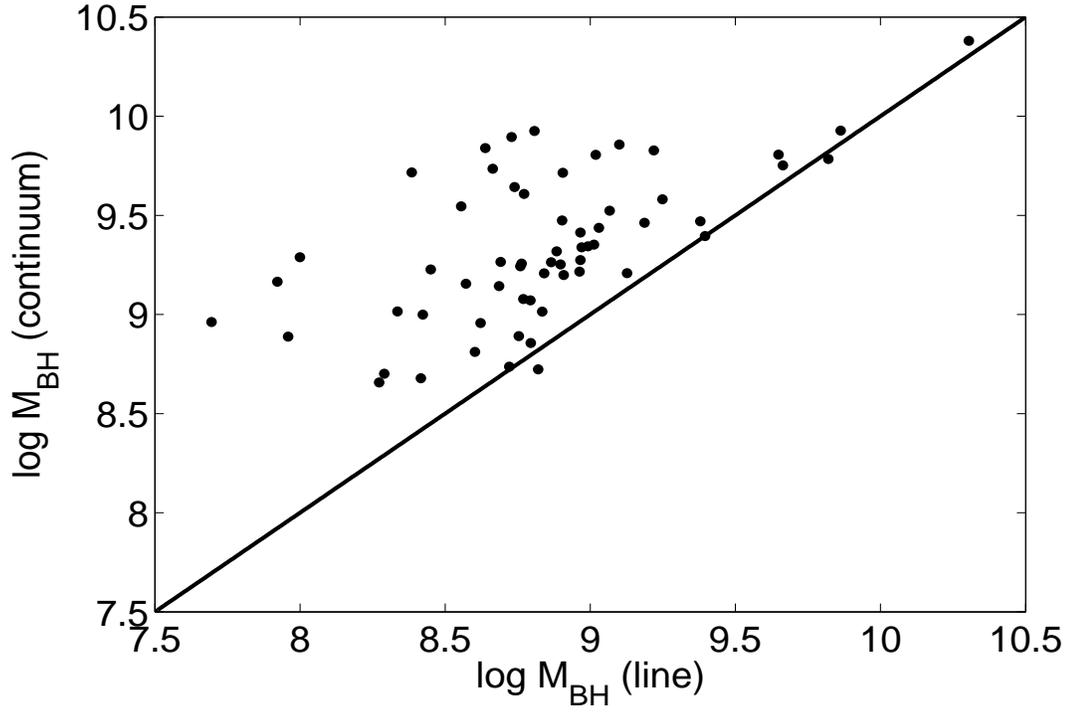,width=150mm,height=100mm,angle=0.0}
\caption{The comparison of the black hole masses estimated with two
different approaches.}
   \label{Fig:lightcurve-ADAri1}
   \end{center}
\end{figure}
\clearpage
\begin{figure}
   \vspace{2mm}
   \begin{center}
   \hspace{3mm}\epsfig{figure=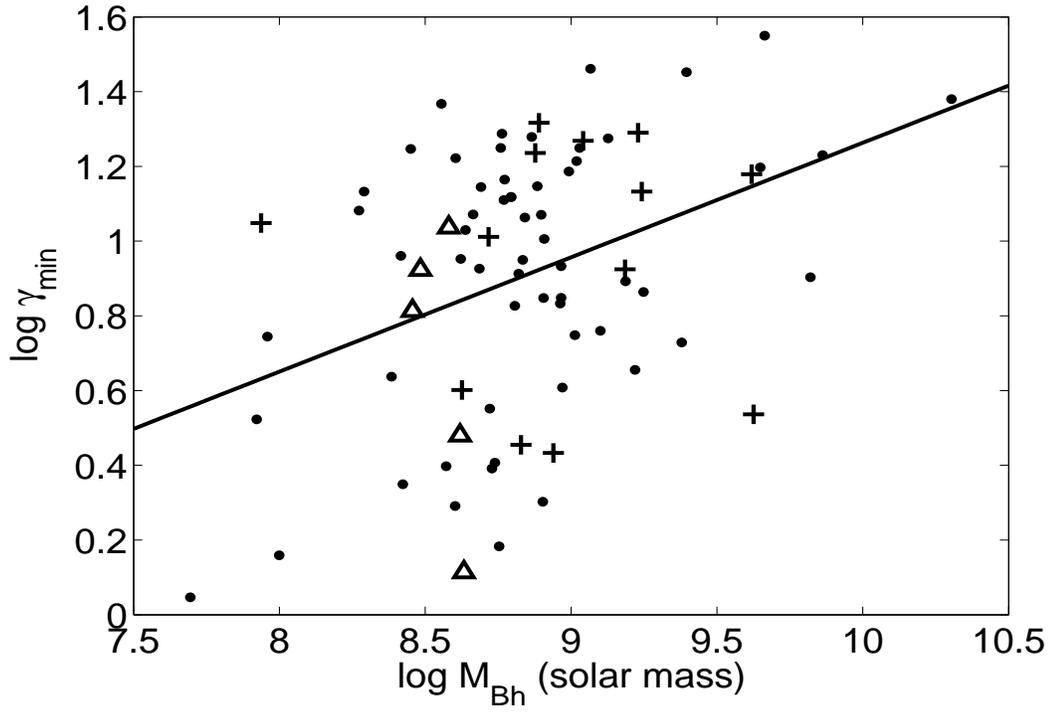,width=150mm,height=100mm,angle=0.0}
\caption{The relation between black hole mass and the minimal
Lorentz factor of the jet. The full circles represent quasars which
$M_{\rm BH}$ estimated by line luminosities, while the triangles
represent BL Lac objects. The crosses represent quasars which
$M_{\rm BH}$ estimated by continuum luminosities.}
   \label{Fig:lightcurve-ADAri1}
   \end{center}
\end{figure}
\clearpage
\begin{figure}
   \vspace{2mm}
   \begin{center}
   \hspace{3mm}\epsfig{figure=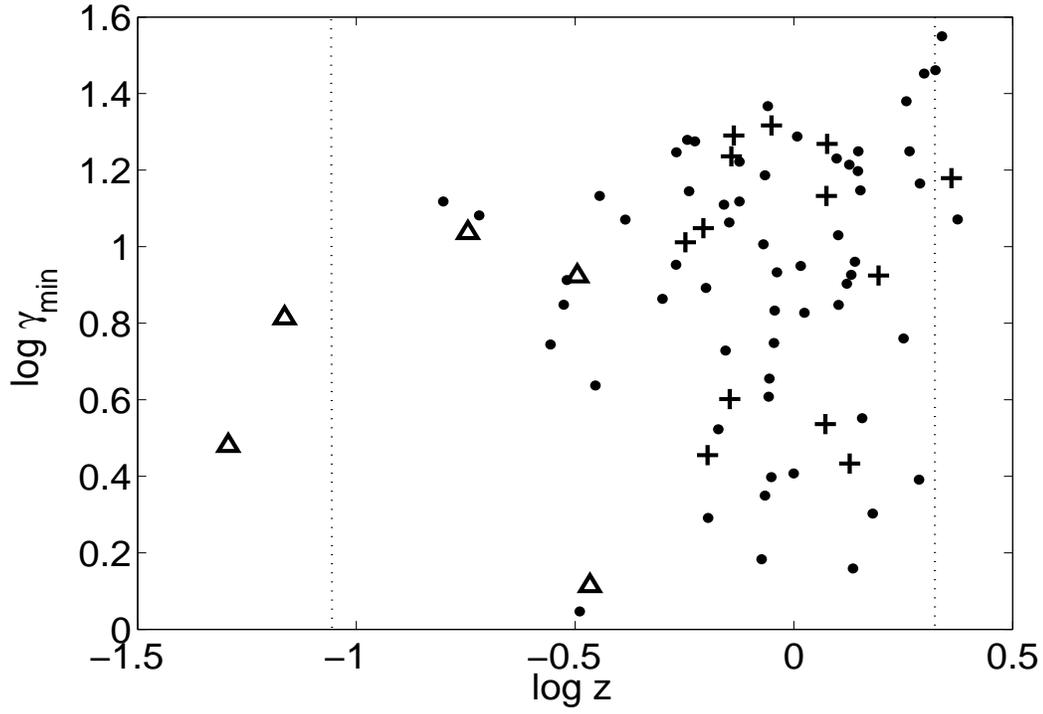,width=150mm,height=100mm,angle=0.0}
\caption{The minimal Lorentz factor of jet versus redshift plane for
our sample (symbols as in Fig. 2). The restricted redshift range,
$0.1<z<2.1$, is indicated by the dotted lines.}
   \label{Fig:lightcurve-ADAri1}
   \end{center}
\end{figure}
\clearpage
\begin{figure}
   \vspace{2mm}
   \begin{center}
   \hspace{3mm}\epsfig{figure=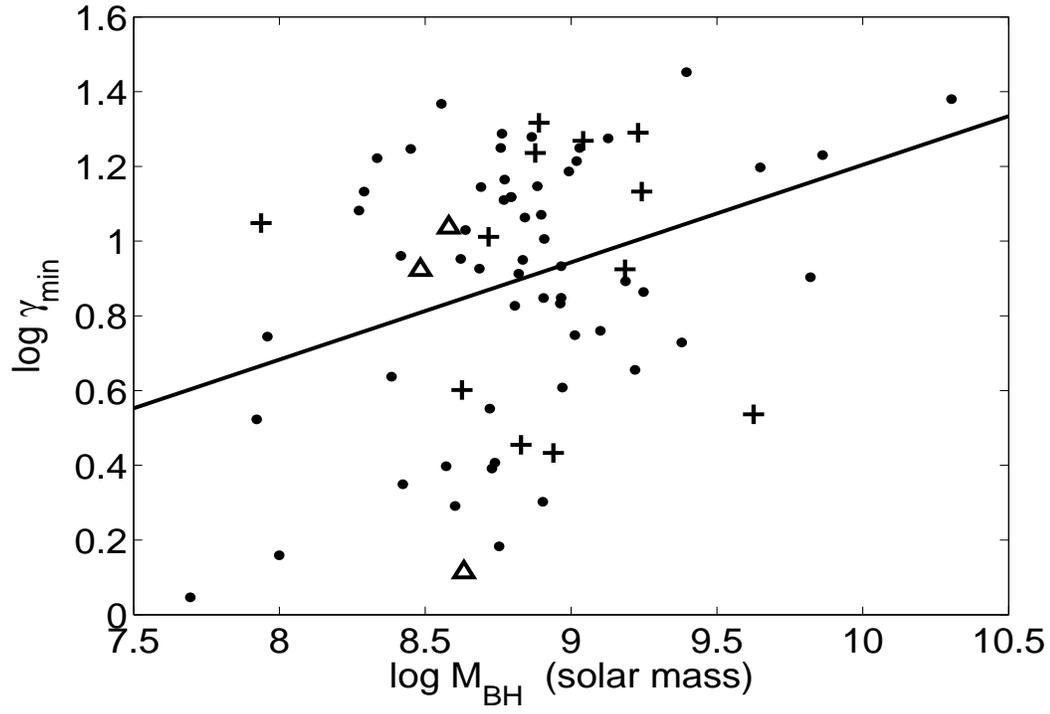,width=150mm,height=100mm,angle=0.0}
   \caption{The same as Fig. 2, but for the subsample within the restricted redshift range.}
   \label{Fig:lightcurve-ADAri1}
   \end{center}
\end{figure}

\clearpage
\begin{figure}
   \vspace{2mm}
   \begin{center}
   \hspace{3mm}\epsfig{figure=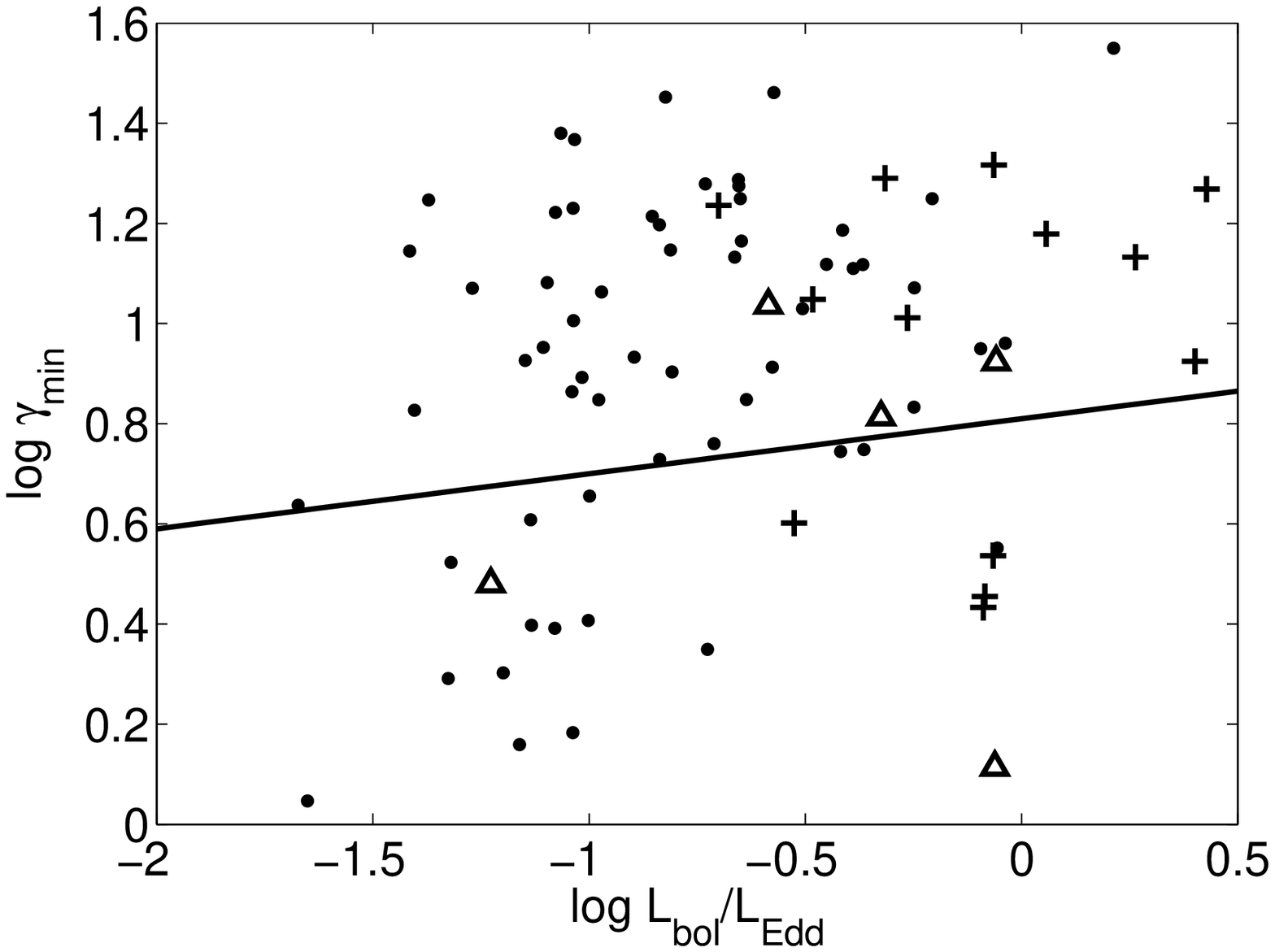,width=150mm,height=100mm,angle=0.0}
\caption{The relation between the Eddington ratio and the minimal
Lorentz factor of the jet (symbols  as in Fig. 2).}
   \label{Fig:lightcurve-ADAri1}
   \end{center}
\end{figure}
\clearpage

\clearpage
\begin{thebibliography}{99}
\bibitem[]{}Baldwin J.A., Wampler E.J., Gaskell C.M., 1989, ApJ, 338, 630 (B89)
\bibitem[]{}Bettoni, D., Falomo, R., Fasano,G., et al., 2003, A\&A, 339, 869
\bibitem[]{}Blandford R. D., Payne D. G., 1982, MNRAS, 199, 883
\bibitem[]{}Blandford R. D., Znajek R. L., 1977, MNRAS, 179, 433
\bibitem[]{}Brotherton M. S. 1996, ApJS, 102, 1 (B96)
\bibitem[]{}Cao X., Jiang D. R., 1999, MNRAS, 307, 802
\bibitem[]{}Cao X., Jiang D. R., 2001, MNRAS, 320, 347
\bibitem[]{}Cao X., Jiang D. R., 2002, MNRAS, 331, 111
\bibitem[]{}Cohen M. H., Lister M. L., Homan D. C., et al., 2007, ApJ, 658, 232
\bibitem[]{}Dibai, \'{E}.~A.\ 1980, Azh., 57, 677 (English translation: Sov.~Astron., 24, 389)
\bibitem[]{}Falcke H., Biermann P., 1995, A\&A, 293, 665
\bibitem[]{}Francis P. J., Hewett P. C., Foltz C. B., et al., 1991, ApJ, 373, 465
\bibitem[]{}Gaskell C. M., Wampler E. J., Shields G. A., 1981, ApJ, 249, 443
\bibitem[]{}Gelderman R., Whittle M. 1994, ApJS, 91, 491 (G94)
\bibitem[]{}Gu M., Cao X., Jiang D. R., 2001, MNRAS, 327, 1111
\bibitem[]{}Henstock D. R., Browne I. W. A., Wilkinson P. N., et al., 1997,MNRAS, 290, 380 (H97)
\bibitem[]{}Ho L. C., Kormendy J., 2000, The Encyclopedia of Astronomy and Astrophysics (Institute of Physics Publishing). (astro-ph/0003267)
\bibitem[]{}Hughes S. A., Blandford Roger D. 2003, ApJ, 585, L101
\bibitem[]{}Hunstead R. W., Murdoch H. S., Shobbrook R. R., 1978, MNRAS, 185, 149(H78)
\bibitem[]{}Jackson N., Browne W. A. 1991, MNRAS, 250, 414 (JB91a)
\bibitem[]{}Jackson N., Browne W. A. 1991, MNRAS, 250, 422 (JB91b)
\bibitem[]{}Kaspi S., Smith P. S., Maoz D., et al., 1996, ApJ, 471, L75
\bibitem[]{}Kaspi S., Smith P. S., Netzer H., et al., 2000, ApJ, 533, 631
\bibitem[]{}Kellermann, K. I., Lister, M. L., D. C. Homan, 2004, ApJ, 609, 539
\bibitem[]{}Laor A. 2000, ApJ, 543, L111
\bibitem[]{}Lawrence C. R., Zucker J. R., REanhead C. S. et al. 1996, ApJS, 107, 541 (L96)
\bibitem[]{}Lister M. L., Homan D. C., 2005, AJ, 130, 1389L
\bibitem[]{}Liu Y., Jiang D. R.,Gu M. F., 2006, ApJ, 637, 669
\bibitem[]{}Macklin J. T., 1982, MNRAS, 199, 1119
\bibitem[]{}Marziani P., Sulentic J. W., Dultzin-Hacyan D. et al. 1996, ApJS, 104, 37 (M96)
\bibitem[]{}McGill K. L., Woo J.-H., Treu T., et al., 2008, ApJ, 673,703
\bibitem[]{}Nerzer H. N., Brotherton M. S., Wills B. J. et al. 1995, ApJ, 448, 27  (N95)
\bibitem[]{}Netzer H., Peterson B.M., 1997, in Astronomical Time Series, eds.,  D. Maoz, A. Sternberg \& E. Leibowitz (Dordrecht: Kluwer), 85
\bibitem[]{}Neugebauer G., Oke J. B., Beckline E. E., Matthews K.,1979, ApJ, 230, 79 (N79)
\bibitem[]{}Oshlack A. Y. K. N., Webster R. L.,  Whiting M. T. 2002, ApJ, 576, 81 (O02)
\bibitem[]{}Peterson B. M., 1993, PASP, 105, 207
\bibitem[]{}Rawlings S. G., Saunders R. D. E., 1991, Nat, 349, 138
\bibitem[]{}Scarpa R., Falomo R., 1997, A\&A, 325, 109 (S97)
\bibitem[]{}Soltan A. 1982, MNRAS, 200, 115
\bibitem[]{}Steidel C. C., Sargent W. L. W. 1991, ApJ, 382, 433 (S91)
\bibitem[]{}Stickel M., Fried W., K\"uhr H., 1989, A\&AS, 80, 103 (S89)
\bibitem[]{}Stickel M., K\"uhr H., 1993, A\&AS, 100, 395 (SK93)
\bibitem[]{}Stickel M., K\"uhr H., Fried J. W. 1993, A \& AS, 97, 483 (S93)
\bibitem[]{}Tadhunter C. N., Morganti R., Alighieri S. S. et al. 1993, MNRAS, 263, 999 (T93)
\bibitem[]{}Urry C. M., Scarpa, R., O'Dowd M., et al., 2000, ApJ, 532, 816 (U00)
\bibitem[]{}Volonteri M., Sikora M., Lasota J.-P.,  2007, ApJ, 667, 704
\bibitem[]{}Wandel A., Peterson B.M., Malkan M.A., 1999, ApJ, 526, 579
\bibitem[]{}Wills B., Browne I. W. A. 1986, ApJ, 302, 56 (W86)
\bibitem[]{}Wills B. J., Thompson K. L., Han M., et al., 1995, ApJ, 447, 139 (W95)
\bibitem[]{}Wu X. B., Wang R., Kong M. Z., et al., 2004, A\&A, 424, 793
\bibitem[]{}Xie G. Z., Dai H., Zhou S. B., 2007, AJ, 134, 1464
\bibitem[]{}Xie Z. H., Hao J. M., Du L. M., Zet al., 2008, PASP, 120, 477
\bibitem[]{}Yu Q., Tremaine S. 2002, MNRAS, 335, 965
\end{thebibliography}
\end{document}